\documentclass[12pt]{article}

\usepackage[english]{babel}
\usepackage{amssymb,latexsym,amsmath,amscd}

\begin{document}

\title
{Yang-Mills like instantons in eight and seven dimensions}
\author
{E.K. Loginov\footnote{{\it E-mail address:} ek.loginov@mail.ru} \,\,and\, E.D. Loginova\footnote{{\it
E-mail address:} loginova.ed@mail.ru}\\}
\date{}
\maketitle

\vspace{-1cm}
\begin{center}
$^{*}${\it Ivanovo State Power Engineering University, Ivanovo, 153003, Russia}\\
$^{\dag}${\it Ivanovo State University, Ivanovo, 153025, Russia}
\end{center}
\vspace{0cm}

\begin{abstract}
We consider a gauge theory in which a nonassociative Moufang loop takes the place of a
structure group. We construct Belavin-Polyakov-Schwartz-Tyupkin (BPST) and t'Hooft like instanton solutions of the gauge theory in
seven and eight dimensions.
\end{abstract}

\section{Introduction}

The pure Yang-Mills theory defined in the four-dimensional Euclidean space has a rich and
interesting structure even at the classical level. The discovery of regular solutions to the
Yang-Mills field equations, which correspond to absolute minimum of the action, see
Ref.~\cite{bela75}, has led to an intensive study of such a classical theory. One hopes that a
deep understanding of the classical theory will be invaluable when one tries to quantize such
a theory. In the past few years, increased attention has been paid to the self-dual gauge
field equations in space-time of dimension greater than four, with a view to obtaining
physically interesting theories via dimensional reduction. Such equations appear in the
many-dimensional theory of supergravity, see Refs.~\cite{engl82,gurs83,duff86}, and in the
low-energy effective string theory, see Refs.~\cite{harv91,guna95,logi05}. Using solutions of
the Yang-Mills equations in $d>4$ makes possible to obtain soliton solutions in these
theories.
\par
From the viewpoint of mathematical physics, some above works has made most conspicuous the
possibly central role played by octonions, see Ref.~\cite{dund84}, and their attending Lie
groups, see Ref.~\cite{faul77}. The algebra of octonions (Cayley numbers) is the most known
example of nonassociative alternative algebras. The alternative algebras are closely
associated with the Malcev algebras and analytic Moufang loops. These algebraic structures are
actively investigated and applied in physics, see Ref.~\cite{okub95}. In Ref.~\cite{logi07},
we extend the gauge invariance to the theory in which a nonassociative Moufang loop takes the
place of the structure group. The present paper is a next step in this direction.
\par
The paper is organized as follows. Section 2 contains well-known facts about the octonions and
mathematical structures connected with one. In Sections 3 and 4, we construct BPST and t'Hooft
like instanton solutions of the nonassociative (octonionic) gauge theory in seven and eight
dimensions respectively.

\section{Preliminaries}

We recall, see Ref.~\cite{scha66}, that the octonionic algebra $\mathbb O$ is a real linear
algebra with the canonical basis $1,e_{1},\dots,e_{7}$ such that
\begin{equation}\label{41-02}
e_{i}e_{j}=-\delta_{ij}+c_{ijk}e_{k},
\end{equation}
where the structure constants $c_{ijk}$ are completely antisymmetric and nonzero and equal to
unity for the seven combinations: $(ijk)=(123)$, (145), (167), (246), (275), (374), (365). The
algebra of octonions is not associative but alternative, i.e. the associator
\begin{equation}\label{41-01}
(x,y,z)=(xy)z-x(yz)
\end{equation}
is totally antisymmetric in $x,y,z$. It is easy to prove that the associator is satisfies the
identities
\begin{equation}\label{41-26}
(x,y,yz)=(x,y,z)y,\qquad (x,y,zy)=y(x,y,z).
\end{equation}
The algebra of octonions is a division algebra. In addition, it is permits the involution
$x\to\bar x$ such that the elements
\begin{equation}
t(x)=x+\bar x\qquad\text{and}\qquad n(x)=\bar xx
\end{equation}
are in $\mathbb R$. In the canonical basis this involution is defined by $\bar e_{i}=-e_{i}$.
Therefore the bilinear form
\begin{equation}\label{41-14}
\langle x,y\rangle=\frac12(\bar xy+\bar yx)
\end{equation}
is positive definite and defines an inner product on $\mathbb O$. It is well
known~\cite{jaco58}, that every automorphism of $\mathbb O$ is inner and the group
$\text{Aut}\,\mathbb O$ is isomorphic to $G_2$. Therefore the inner product (\ref{41-14}) is
invariant under all automorphisms of $\mathbb O$, i.e.
\begin{equation}
\langle Ux,Uy\rangle=\langle x,y\rangle
\end{equation}
for any $u\in\text{Aut}\,\mathbb O$. Besides, it can be proved that the quadratic form $n(x)$
permits the composition
\begin{equation}\label{41-06}
n(xy)=n(x)n(y).
\end{equation}
It follows from here that
\begin{equation}\label{41-20}
\langle [x,y],z\rangle=\langle x,[y,z]\rangle.
\end{equation}
\par
Since the algebra of octonions is nonassociative, its commutator algebra $\mathbb O^{(-)}$ is
non-Lie. Instead of the Jacobi identity it satisfies the identity
\begin{equation}\label{41-22}
[[x,y],z]+[[y,z],x]+[[z,x],y]=6(x,y,z),
\end{equation}
which is sometimes called the generalized Jacobi identity (see, for example~\cite{akiv06}).
Let us consider the subalgebra of $\mathbb O^{(-)}$ given by
\begin{equation}\label{41-07}
\mathbb M=\{x\in\mathbb O^{(-)}\mid t(x)=0\}.
\end{equation}
Obviously, $\mathbb M$ is a seven-dimensional non-Lie algebra. It satisfies the Malcev
identity
\begin{equation}
J(x,y,[x,z])=[J(x,y,z),x],
\end{equation}
where $J(x,y,z)=[[x,y],z]+[[y,z],x]+[[z,x],y]$ is the Jacobian of the elements $x,y$ and $z$,
and therefore it is called the Malcev algebra. The algebra $\mathbb M$ has the basis
$e_{1},\dots,e_{7}$. Using Eq.~(\ref{41-02}) we can find the commutators and associators of
the basis elements
\begin{align}
[e_{i},e_{j}]&=2c_{ijk}e_{k}\label{41-27},\\
(e_{i},e_{j},e_{k})&=2c_{ijkl}e_{l}.\label{41-17}
\end{align}
where $c_{ijkl}$ is a completely antisymmetric nonzero tensor equal to unity for the seven
combinations $(ijkl)=(4567)$, (2367), (2345), (1357), (1364), (1265), (1274). It can easily be
checked that $c_{ijk}$ and $c_{ijkl}$ satisfy the identities
\begin{align}
c_{mni}c_{psi}&=\delta_{mp}\delta_{ns}-\delta_{ms}\delta_{np}+c_{mnps},\label{41-23}\\
c_{mnij}c_{pij}&=4c_{mnp},\label{41-25}\\
c_{mnij}c_{psij}&=4\delta_{mp}\delta_{ns}-4\delta_{ms}\delta_{np}+2c_{mnps}.\label{41-21}
\end{align}
Finally, since the quadratic form $n(x)$ permits the composition (\ref{41-06}), the set
\begin{equation}\label{41-08}
\mathbb S=\{x\in\mathbb O\mid n(x)=1\}
\end{equation}
is closed relative to the multiplication in $\mathbb O$ and hence it is an analytic loop. The
loop $\mathbb S$ is unique, to within an isomorphism, analytic compact simple nonassociative
Moufang loop. Its tangent algebra is isomorphic to the Malcev algebra (\ref{41-07}).

\section{Octonionic instantons in seven dimensions}

Let $A_{m}(x)$ be a vector field defined in $\mathbb R^7$ and taking its values in the Malcev
algebra $\mathbb M$. We fixes the canonical basis $e_{1},\dots,e_{7}$ in $\mathbb M$ and
consider the action
\begin{equation}\label{43-01}
S=\int[\langle F_{mn},F_{mn}\rangle+\langle J_{m},A_{m}\rangle]d^7x,
\end{equation}
where $A_{m}(x)=A_{m}^{i}(x)e_{i}$ and the field strength tensor $F_{mn}(x)$ and the vector
field $J_{m}(x)$ are defined by
\begin{align}
F_{mn}&=\partial_{m}A_{n}-\partial_{n}A_{m}+[A_{m},A_{n}],\\
J_{m}&=\frac12c_{mnps}(A_{n},A_{p},A_{s}).
\end{align}
Here $(A_{n},A_{p},A_{s})$ is the associator of the vectors fields and $c_{mnps}$ is the fully
antisymmetric tensor defined in by Eq.~(\ref{41-17}). Using (\ref{41-20}) and (\ref{41-21}), we
rewrite the Lagrangian in~(\ref{43-01}) by
\begin{equation}\label{43-27}
\mathcal L=\langle F_{mn},F_{mn}\rangle-\frac14c_{mnps}\langle
F_{mn},F_{ps}\rangle+\partial_{m}I_{m},
\end{equation}
where the vector field
\begin{equation}
I_{m}=c_{mnps}\left[\langle A_{n},\partial_{p}A_{s}\rangle+\frac23\langle
A_{n,}A_{p}A_{s}\rangle\right].
\end{equation}
Since the inner product (\ref{41-14}) is invariant under all automorphisms of $\mathbb O$ and
the variation of field on the boundary of volume vanish in the deduction of equations of
motion, if follows that the Lagrangian is invariant with respect to the transformation
\begin{equation}\label{42-20}
F_{mn}\to UF_{mn},
\end{equation}
where $U(x)$ is  a function taking its values in the group of all automorphisms of $\mathbb
M$. From this it follows that the functional (\ref{43-01}) is invariant with respect to the
gauge transformation that was defined in Ref.~\cite{logi07}.
\par
Substituting the obvious equality $\langle e_i,e_j\rangle=2\delta_{ij}$ in the action
(\ref{43-01}), we easily get the equations of motion
\begin{equation}\label{43-02}
[D_{m},F_{mn}]=J_{n}.
\end{equation}
where the covariant operator $D_{n}=\partial_{n}+A_{n}$. Using the identity (\ref{41-22}), we
prove the Bianchi identity for Moufang gauge fields in seven dimensions
\begin{equation}
c_{mnps}[D_{n},F_{ps}]+4J_{m}=0.
\end{equation}
It follows that every solution of the equations
\begin{equation}\label{43-03}
F_{mn}=\frac14c_{mnps}F_{ps}
\end{equation}
is a solution of the equations of motion (\ref{43-02}). Since
\begin{equation}
\langle F_{mn},F_{mn}\rangle-\frac14c_{mnps}\langle F_{mn},F_{ps}\rangle=
\frac23\left|F_{mn}-\frac14c_{mnps}F_{ps}\right|^2\geqslant0,
\end{equation}
it follows that such solutions in fact give the absolute minimum of the action (\ref{43-01}).
\par
Now we turn to a search of solutions of the self-dual equations (\ref{43-03}). We will use the
following construction, see Ref.~\cite{logi07}. Let $u$ be a fixed element of $\mathbb S$. We
define a new multiplication in $\mathbb O$ by
\begin{equation}\label{43-04}
x\circ y=(xu^{-3})(u^{3}y).
\end{equation}
This multiplication converts the vector space $\mathbb O$ into a linear algebra. We denote
this algebra by the symbol $\mathbb O'$. It is easy to prove that the algebras $\mathbb O$ and
$\mathbb O'$ are isomorphic. On the other hand, this isomorphism induces the isomorphism
$\mathbb O^{(-)}\simeq\mathbb O'^{(-)}$ of their commutator algebras and hence the isomorphism
$\mathbb M\simeq\mathbb M'$, where
\begin{equation}\label{43-05}
\mathbb M'=\{x\in\mathbb O'^{(-)}\mid t(x)=0\}.
\end{equation}
From this it follows that $\mathbb M'$ also is the seven-dimension Malcev algebras. We suppose
that $e_{1},\dots,e_{7}$ is a standard basis of $\mathbb M$ and $y(x)=y_{k}(x)e_{k}$, where
$y_{k}(x)$ are real-valued functions. Since the gauge field $A_{m}$ and the field strength
tensor $F_{mn}$ take its values in $\mathbb M$, the isomorphism $\mathbb M\to\mathbb M'$
induces the transformations $A_{m}\to A'_{m}$ and
\begin{equation}\label{43-07}
F_{mn}\to F'_{mn}=\partial_{[m}A'_{n]}+A'_{[m}u^{-3}\cdot u^{3}A'_{n]},
\end{equation}
where $u(x)$ is a field taking its values in the loop $\mathbb S$. We select this field so
that
\begin{equation}\label{43-23}
y=u^{3}|y|.
\end{equation}
Now we choose the ansatz
\begin{equation}\label{43-06}
A'_{m}=\frac12[y,e_{m}].
\end{equation}
Substituting this ansatz in Eq.~(\ref{43-07}) and then using the identities (\ref{41-26}), we
get
\begin{align}
F'_{mn}&=\partial_{[m}A'_{n]}+A'_{[m}A'_{n]}+y^{-1}(y,A'_{[m},A'_{n]})\nonumber\\
&=[\partial_{[m}y,e_{n]}]+[y,e_{[m}][y,e_{n]}]-(y,[y,e_{[m}],e_{n]}).
\end{align}
Using Eqs.~(\ref{41-27}) and (\ref{41-17}), we rewrite the obtained expression for the field
strength tensor in the form
\begin{equation}
F'_{mn}=\{c_{pk[n}\partial_{m]}y_{p}+2(c_{pim}c_{nsj}c_{ijk}
+c_{pi[m}c_{n]isk})y_{p}y_{s}\}e_{k}.
\end{equation}
In order that to simplify this expression we note that
\begin{align}
c_{mni}c_{psj}c_{ijk}&=\delta_{s[m}c_{n]pk}-\delta_{p[m}c_{n]sk}
+c_{mni}c_{ipsk}+c_{psi}c_{imnk},\label{41-28}\\
c_{mni}c_{msj}&=\delta_{ij}\delta_{ns}-c_{ijk}c_{nsk}-2c_{ijns},\label{41-29}\\
c_{mni}c_{mnj}&=7\delta_{ij}.\label{41-30}
\end{align}
Then using Eqs.~(\ref{41-23}) and (\ref{41-28}), we get
\begin{equation}\label{43-09}
F'_{mn}=(c_{pk[n}\partial_{m]}y_{p}-2c_{pk[n}y_{m]}y_{p}-2c_{mnk}y_{p}y_{p})e_{k}.
\end{equation}
Now we select the function $y=y(x)$ in the form
\begin{equation}\label{43-08}
y=-\frac{x-b}{\lambda^2+|x-b|^2},
\end{equation}
there $b$ and $\lambda$ are eight arbitrary constant parameters. Substituting
Eq.~(\ref{43-08}) in Eq.~(\ref{43-09}), we obtain
\begin{equation}\label{43-10}
F'_{mn}=\frac{2c_{mnk}e_{k}}{(\lambda^2+|x-b|^2)^2}.
\end{equation}
By the identity (\ref{41-25}), it follows that the tensor (\ref{43-10}) is self-dual.
Obviously, this is a BPST like instanton solutions in seven dimensions. However in spite of
the fact that the field strength tensor (\ref{43-10}) fall of as $1/x^4$, the functional
(\ref{43-01}) diverges. The point is that the term $\partial_{m}I_{m}$ in the right hand side
of Eq.~(\ref{43-27}) fall of only as $1/x^6$. Thus this requirement is not a useful way to
impose boundary conditions on the solutions we are considering. Note also that similar
problems arise in the Yang-Mills theory. It is known, see Ref.~\cite{jaff80} that there are no
finite action solutions to the Yang-Mills equations if $d>4$.
\par
Now we will follow a method proposed by t'Hooft~\cite{hoof76} and developed by several
authors~(see e.g.~\cite{raja82}). The t'Hooft like solutions is obtained if we lay on the
field strength tensor (\ref{43-09}) the condition
\begin{equation}\label{43-13}
c_{mns}F'_{mn}=0.
\end{equation}
It follows from Eq.~(\ref{41-23}) that this condition is equivalent to the requirement of
antiself-duality of $F'_{mn}$. Using Eqs.~(\ref{41-29}) and (\ref{41-30}), we prove that
\begin{equation}\label{43-11}
c_{mns}F'_{mn}=-2(\partial_{m}y_{m}+5y_{m}y_{m})e_{s}.
\end{equation}
A solution of Eq.~(\ref{43-11}) we look for in the form
\begin{equation}\label{43-12}
y_{m}=k\frac{\partial_{m}\varphi}{\varphi},
\end{equation}
where $\varphi$ is a function of $x_{i}$. It is easily be checked that the condition
(\ref{43-13}) is true if $k=1/5$ and
\begin{equation}
\partial_{m}\partial_{m}\varphi=0.\label{43-14}
\end{equation}
We choose the follows solution of Eq.~(\ref{43-14}):
\begin{equation}\label{43-31}
\varphi=1+\sum^{N}_{i=1}\frac{\lambda^2_{i}}{|x-b_{i}|^5},
\end{equation}
where $\lambda_{i}$ and $b_{i}$ are arbitrary constant parameters. In this case the the gauge
field $A'_{\mu}$ fall of as $1/x^6$. At the same time the function (\ref{43-31}) becomes
singular as $x\to b_{i}$. Note that BPST like octonionic solution in seven dimensions similar
to (\ref{43-10}) was obtained in Ref.~\cite{logi07a}.

\section{Octonionic instantons in eight dimensions}

In order to extend the found above solutions to eight dimensions we must use the following
construction. Let $f_{mnps}$ is a fully antisymmetric tensor such that
\begin{equation}\label{43-22}
\begin{aligned}
f_{ijk0}&=c_{ijk},\\
f_{ijkl}&=c_{ijkl},
\end{aligned}
\end{equation}
where the tensors $c_{ijk}$ and $c_{ijkl}$ was introduced above. We define an analog of the
t'Hooft tensor in eight dimensions by
\begin{equation}\label{43-15}
f_{mn}=\frac12(\bar e_{m}e_{n}-\bar e_{n}e_{m}),
\end{equation}
where $e_{m}$ is a basis element of the octonionic algebra. It is easy to prove that the
tensor
\begin{equation}\label{43-16}
f_{mn}=f_{mn}^{k}e_{k},
\end{equation}
where the coefficients
\begin{equation}
f_{mn}^{k}=\delta_{m0}\delta_{nk}-\delta_{n0}\delta_{mk}-c_{mnk}.
\end{equation}
Using Eq.~(\ref{43-16}), we prove that the tensor (\ref{43-15}) is self-dual. It satisfies the
identity
\begin{equation}\label{43-33}
f_{mnps}f_{ps}=6f_{mn}.
\end{equation}
Besides, the following analogs of the identities (\ref{41-23}) and
(\ref{41-28})--(\ref{41-30}) are true:
\begin{align}
f_{mn}^{i}f_{ps}^{j}c_{ijk}&=\delta_{s[m}f_{n]p}^{k}-\delta_{p[m}f_{n]s}^{k}
+f_{mn}^{i}c_{ipsk}+f_{ps}^{i}c_{imnk},\label{43-20}\\
f_{mn}^{i}f_{ps}^{i}&=\delta_{mp}\delta_{ns}-\delta_{ms}\delta_{np}+f_{mnps},\label{43-21}\\
f_{mn}^{i}f_{ms}^{j}&=\delta_{ij}\delta_{ns}-c_{ijk}f_{ns}^{k}-2c_{ijns},\label{43-25}\\
f_{mn}^{i}f_{mn}^{j}&=8\delta_{ij}.\label{43-26}
\end{align}
\par
Now we can construct (anti)self-dual solutions in eight dimensions. Let $A_{m}(x)$ be a vector
field defined in $\mathbb R^8$ and taking its values in the Malcev algebra $\mathbb M$. We
fixes the canonical basis $e_{0}=1,e_{1},\dots,e_{7}$ in $\mathbb O$ and consider the action
We consider the action
\begin{equation}\label{43-17}
S=\int[\langle F_{mn},F_{mn}\rangle+\langle J_{m},A_{m}\rangle]d^8x,
\end{equation}
where $A_{m}(x)=A_{m}^{a}(x)e_{a}$ and the field strength tensor $F_{mn}(x)$ and the vector
field $J_{m}(x)$ are defined by
\begin{align}
F_{mn}&=\partial_{m}A_{n}-\partial_{n}A_{m}+[A_{m},A_{n}],\label{42-14}\\
J_{m}&=\frac13f_{mnps}(A_{n},A_{p},A_{s}).
\end{align}
As above, we prove that the functional (\ref{43-17}) is gauge invariant and the corresponding
equations of motion have the form
\begin{equation}\label{43-18}
[D_{m},F_{mn}]=J_{n}.
\end{equation}
Using the identity (\ref{41-22}), we prove the Bianchi identity for Moufang gauge fields in
eight dimensions
\begin{equation}
f_{mnps}[D_{n},F_{ps}]+6J_{m}=0.
\end{equation}
It follows that every solution of the equations
\begin{equation}\label{43-19}
F_{mn}=\frac16f_{mnps}F_{ps}
\end{equation}
is a solution of the equations of motion (\ref{43-18}). By the argument as above, we see that
such solutions give the absolute minimum of the action (\ref{43-17}).
\par
Now we turn to a search of solutions of the self-dual equations (\ref{43-19}). Let
$\{e_0=1,e_1\dots,e_7\}$ be the canonical basis of $\mathbb O$ and let the function
$y(x)=y_{s}(x)e_{s}$ satisfies Eq.~(28). We choose the ansatz
\begin{equation}\label{43-34}
A'_{m}=f_{mn}y_{n}.
\end{equation}
A straightforward calculation shows that $f_{ij}=-c_{ijk}e_{k}$ and $f_{0j}=e_{j}$ as
$i,j\ne0$. Since $y_{n}(x)$ are real-valued functions, it follows that the field $A'_{m}(x)$
takes its values in $\mathbb M'$. Now we find the field strength tensor
\begin{equation}\label{43-29}
F'_{mn}=\{f^{k}_{p[m}\partial_{n]}y_{p}+2(f_{mp}^{i}f_{ns}^{j}c_{ijk}
+f^{i}_{p[m}c_{n]sik})y_{p}y_{s}\}e_{k}.
\end{equation}
In order that to simplify this expression we use the identities (\ref{43-20}) and
(\ref{43-21}). As a result, we obtain the field strength tensor
\begin{equation}\label{43-24}
F'_{mn}=(f_{p[n}^{k}\partial_{m]}y_{p}-2f_{p[n}^{k}y_{m]}y_{p}-2f_{mn}^{k}y^2)e_{k}.
\end{equation}
We select the function $y=y(x)$ in the form
\begin{equation}\label{43-28}
y=-\frac{x-b}{\lambda^2+|x-b|^2},
\end{equation}
where $b$ and $\lambda$ are nine arbitrary constant parameters. Substituting Eq.~(\ref{43-28})
in Eq.~(\ref{43-29}), we obtain
\begin{equation}\label{43-30}
F'_{mn}=\frac{2f^{k}_{mn}e_{k}}{(\lambda^2+|x-b|^2)^2}.
\end{equation}
By the identity (\ref{43-33}), it follows that the tensor (\ref{43-30}) is self-dual.
Therefore this is a BPST like instanton solutions in eight dimensions. Arguing as above, we
see that the functional (\ref{43-17}) also diverges. Note that the field strength tensor of
the form (\ref{43-30}) was previously met in~\cite{oots05}. However, the equation of motion
for it, (i.e., the equation (\ref{43-18})) were not found.
\par
Now we lay on the field strength tensor (\ref{43-24}) the condition
\begin{equation}
f_{mn}^{s}F'_{mn}=0.
\end{equation}
It follows from Eq.~(\ref{43-21}) that this condition is equivalent to the requirement of
antiself-duality of $F'_{mn}$. Using Eqs.~(\ref{43-25}) and (\ref{43-26}), we prove that
\begin{equation}
f_{mn}^{s}F'_{mn}=-2(\partial_{m}y_{m}+6y_{m}y_{m})e_{s}.
\end{equation}
A solution of the obtained equation we look for in the form (\ref{43-12}) with $k=1/6$.
Further, we lay on the function $\varphi(x)$ the condition (\ref{43-14}) and find its solution
\begin{equation}\label{43-32}
\varphi=1+\sum^{N}_{i=1}\frac{\lambda^2_{i}}{|x-b_{i}|^6},
\end{equation}
As result, we get the t'Hooft like solution of the equations of motion. In this case the
potential $A'_{\mu}$ fall of as $1/x^7$ and the function (\ref{43-32}) is singular.

\section{Conclusion and Discussion}

In this paper, we have considered a gauge theory in which a nonassociative Moufang loop takes
the place of a structure group. We have constructed BPST and t'Hooft like instanton solutions
of the octonionic gauge theory in seven and eight dimensions.
\par
Note that octonionic instanton solutions in seven and eight dimensions previously were
studied. The BPST like instanton solution were found in Refs.~\cite{fair84,fubi85,ivan92}. The
t'Hooft like multi-instanton solution was found in Ref.~\cite{logi05a}. However all these
solutions are solutions of the standard Yang-Mills field equations and hence they are
fundamentally different from the solutions that were found in the paper. Indeed, the field
strength tensor $F'_{mn}(x)$ in (\ref{43-07}) is defined not only by the vector field
$A_{m}(x)$ but also by the scalar field $u(x)$. As shown in Ref.~\cite{logi07}, the field
$u(x)$ is not removed by the gauge transformations and therefore it is a real field of the
model. Whereas in the Yang-Mills case such scalar fields are absent. The same can be said
about the found instanton solutions. It follows from the condition (\ref{43-23}) that these
solutions are depend on  the choice of the scalar field.
\par
In this regard, the question arises about the physical sense of this field. We risk to suppose
that the field is related to the Higgs field. But of course this is only a hypothesis.

\end{document}